\documentclass[twocolumn,showpacs,preprintnumbers,amsmath,amssymb]{revtex4}
\usepackage{graphicx}
\usepackage{dcolumn}
\usepackage{bm}

\begin{document}

\preprint{APS/123-QED}
\title{General stability criterion of two-dimensional inviscid parallel flow}
\author{Liang Sun}
\email{sunl@mail.ustc.edu.cn;sunl@ustc.edu}
\affiliation{Dept. of Modern Mechanics, \\
 University of Science and Technology of China, Hefei, 230027, P.R.China.}

\date{\today}
\begin{abstract}
 General stability criterions of two-dimensional inviscid parallel flow are
obtained analytically for the first time. First, a criterion for
stability is found as $\frac{U''}{U-U_s}>-\mu_1$ everywhere in the
flow, where $U_s$ is the velocity at inflection point, $\mu_1$ is
eigenvalue of Poincar\'{e}'s problem. Second, we also prove a
principle that the flow is stable, if and only if all the
disturbances with $c_r=U_s$ are neutrally stable.  Finally,
following this principle, a criterion for instability is found as
$\frac{U''}{U-U_s}<-\mu_1$ everywhere in the flow. These results
extend the former theorems obtained by Rayleigh, Tollmien and Fj\o
rtoft and will lead future works to investigate the mechanism of
hydrodynamic instability.

\end{abstract}
\pacs{47.15.Fe, 47.15.Ki, 47.20.Cq, 47.20.-k} \maketitle

The stability of flow is one of fundamental and the most
attracting problems in many fields, such as fluid dynamics,
astrodynamics, oceanography, meteorology et al. The general
stability criterion of inviscid parallel flow is very important
for both theoretic research and application. A general way to
study the problem is investigating the growth of linear
perturbations by means of normal mode expansion, which leads to
famous Rayleigh's equation \cite{Rayleigh1880}. Using this
equation, Rayleigh \cite{Rayleigh1880} first proved the Inflection
Point Theorem, which is a necessary criterion for instability.
Later, Fj{\o}rtoft\cite{Fjortoft1950} found another necessary
criterion for instability, which is a stronger version of
Rayleigh's Theorem. These theorems are well known and are applied
to understand the mechanism of hydrodynamics instability
\cite{Drazin1981,Huerre1998,CriminaleBook2003}. Unfortunately,
both theorems are necessary criterions for instability, neither of
them can give a sufficient condition for the instability. And a
sufficient condition for the stability is also needed for
applications. The aim of this letter is to find a criterion for
these flows.

For this purpose, Rayleigh's equation for an inviscid parallel
flow is employed
\cite{Drazin1981,Huerre1998,SchmidBook2000,CriminaleBook2003}. For
a parallel flow with mean velocity $U(y)$, the streamfunction of
the disturbance expands as series of waves (normal modes) with
real wavenumber $k$ and complex frequency
$\omega=\omega_r+i\omega_i$, where $\omega_i$ relates to the grow
rate of the waves. The flow is unstable if and only if
$\omega_i>0$. We study the stability of the disturbances by
investigating the growth rate of the waves, this method is known
as normal mode method. The amplitude of waves, namely $\phi$,
holds
 \begin{equation}
 (U-c)(\phi''-k^2 \phi)-U''\phi=0,
 \label{Eq:stable_paralleflow_RayleighEq}
 \end{equation}
where $c=\omega/k=c_r+ic_i$ is the complex phase speed. The real
part of complex phase speed $c_r=\omega_r/k$ is the wave phase
speed. This equation is to be solved subject to homogeneous
boundary conditions
\begin{equation}
\phi=0 \,\, at\,\, y=a,b.
\end{equation}
It is obvious that the criterion for stability is $\omega_i=0$
($c_i=0$), for the complex conjugate quantities $\phi^*$ and $c^*$
are also the physical solution of
Eq.(\ref{Eq:stable_paralleflow_RayleighEq}).

 From Rayleigh's
equation, we get the following equation
\begin{equation}
\begin{array}{l}
\displaystyle\int_{a}^{b}
[(\|\phi'\|^2+k^2\|\phi\|^2)+\frac{U''(U-c_r)}{\|U-c\|^2}\|\phi\|^2]\,
dy=0\\
\displaystyle c_i\int_{a}^{b} \frac{U''}{\|U-c\|^2}\|\phi\|^2\,
dy=0. \label{Eq:stable_parallelflow_Rayleigh_Int}
 \end{array}
 \end{equation}
Rayleigh and Fj\o rtoft proved their theorems by
Eq.(\ref{Eq:stable_parallelflow_Rayleigh_Int}), which still plays
an important role in the following discussion. To find a stronger
criterion, we need estimate the rate of $\int_{a}^{b} \|\phi'\|^2
dy$ to $\int_{a}^{b} \|\phi\|^2 dy$. This is known as
Poincar\'{e}'s problem:
\begin{equation}
\int_{a}^{b}\|\phi'\|^2 dy=\mu\int_{a}^{b}\|\phi\|^2 dy,
\label{Eq:stable_paralleflow_Poincare}
\end{equation}
where the eigenvalue $\mu$ is  positive definition for $\phi \neq
0$. The smallest eigenvalue value, namely $\mu_1$, can be
estimated as $\mu_1>(\frac{\pi}{b-a})^2$.

Then there is a criterion for stability using Poincar\'{e}'s
relation (\ref{Eq:stable_paralleflow_Poincare}), a new stability
criterion may be found: the flow is stable if
$\frac{U''}{U-U_s}>-\mu_1$ everywhere.

To get this criterion, we introduce an auxiliary function
$f(y)=\frac{U''}{U-U_s}$,
%
%
%
where $f(y)$ is finite at inflection point. We will prove the
criterion by two steps. At first, we prove result 1: if the
velocity profile is subject to $f(y)>-\mu_1$, then $c_r$ can not
be $U_s$.

Proof: Otherwise,
\begin{equation}
   -\mu_1<\frac{U''}{U-U_s}=\frac{U''(U-U_s)}{(U-U_s)^2}\leq\frac{U''(U-U_s)}{(U-U_s)^2+c_i^2},
\end{equation}
and if $c_r=U_s$, this yields to
\begin{equation}
\begin{array}{rl} \displaystyle\int_a^b
[(\|\phi'\|^2+k^2\|\phi\|^2)+\frac{U''(U-U_s)}{\|U-c\|^2}\|\phi\|^2]\,
dy &\geq \\
\displaystyle\int_a^b [(\mu_1+\frac{U''(U-U_s)}{\|U-c\|^2})
\|\phi\|^2]&>0.

\end{array}
\end{equation}
This contradicts Eq.(\ref{Eq:stable_parallelflow_Rayleigh_Int}).
So result 1 is proved.

Then, we prove result 2: if $-\mu_1<f(y)$ and $c_r\neq U_s$, there
must be $c_i^2=0$.

Proof: Otherwise if $c_i^2\neq0$, so according to
Eq.(\ref{Eq:stable_parallelflow_Rayleigh_Int}), for any arbitrary
number $U_t\in \mathbb{R}$  which does not depend on y, it holds
\begin{equation}
\displaystyle\int_a^b
[(\|\phi'\|^2+k^2\|\phi\|^2)+\frac{U''(U-U_t)}{\|U-c\|^2}\|\phi\|^2]\,
dy=0.
\label{Eq:stable_paralleflow_Sun_Int} \end{equation}
But the above Eq.(\ref{Eq:stable_paralleflow_Sun_Int}) can not be
hold for some special $U_t$. For example, let $U_t=2c_r-U_s$,
then there is $(U-U_s)(U-U_t)<\|U-c\|^2$, and
 \begin{equation}
\frac{U''(U-U_t)}{\|U-c\|^2}=
f(y)\frac{(U-U_s)(U-U_t)}{\|U-c\|^2}>-\mu_1.
\label{Eq:stable_paralleflow_Sun_Ust}
 \end{equation}
This yields to
\begin{equation} \int_a^b
\{\|\phi'\|^2+[k^2+\frac{U''(U-U_t)}{\|U-c\|^2}]\|\phi\|^2\} dy>0,
\end{equation}
which also contradicts Eq.(\ref{Eq:stable_paralleflow_Sun_Int}).
So the second result is also proved.

\begin{figure}
  \includegraphics[width=6cm]{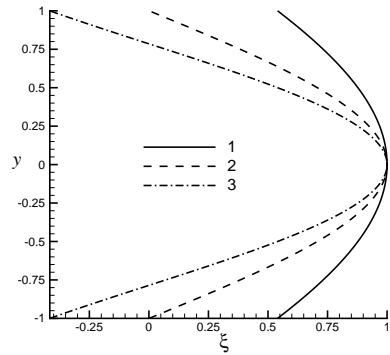}
\caption{vorticity profiles, profile 2 ( $\xi=\cos(\pi y/2)$,
dashed) is neutrally stable, while profile 1 ($\xi=\cos( y)$,
solid) and profile 3 ($\xi=\cos(2y)$, dash doted) are stable and
unstable, respectively.} \label{Fig:vorticity_profile}
\end{figure}

Using 'result 1: if $f(y) > -\mu_1$ then $c_r\neq U_s$' and
'result 2: if $f(y) > -\mu_1$ and $c_r\neq U_s$ then $c_i = 0$',
we find a stability criterion. Theorem 1: If the velocity profile
satisfy $\frac{U''}{U-U_s}>-\mu_1$ everywhere in the flow, it is
stable. This criterion is more powerful than Fj\o rtoft's Theorem.
As known from Fj\o rtoft's theorem, the necessary condition for
instability is that the base vorticity $\xi=U'$ has a local
maximal in the profile. Noting $U''/(U-U_s)\approx \xi_s''/\xi_s$
near the inflection point, where $\xi_s$ is the vortictiy at
inflection point, it means the base vorticity $\xi$ must be convex
enough near the local maximum for instability. As shown in
Fig.\ref{Fig:vorticity_profile}, there are three vorticity
profiles, which have local maximal at $y=0$. Profile 2
($U=\sin(\pi y/2)$) is neutrally stable, while profile 1
($U=\sin(y)$) and profile 3 ($U=\sin(2y)$) are stable and
unstable, respectively.

Both Theorem 1 and Fig.\ref{Fig:vorticity_profile} show that it is
the vorticity profile rather than the velocity profile that
dominates the stability of the flow. This means that the
distribution of vorticity dominates the shear instability in
parallel inviscid flow, which is of essence for understanding the
role of vorticity in fluid. In fact, we can control the
hydrodynamic instability just by controlling the vorticity
distribution according these results. This is an very fascinating
problem, but can not be discussed here. These results may shed
light on the investigation of vortex dynamics.

If the flow is unstable, there are some unstable disturbances with
positive growth rate. For each wavenumber $k$, the fastest growing
disturbance is the most important one. So, what the real phase
velocity $c_r$ is for the fastest growing disturbance? Here we
will show that the real phase velocity $c_r$ of the fastest
growing disturbance is just the velocity at inflection point
$U_s$.

Result 3: If the velocity profile is subject to
$-p\mu_1<f(y)<-\mu_1$ with $1<p<\infty$, then fastest-growing
disturbance with highest $c_i$ must have phase velocity $c_r=U_s$.

Proof: Suppose $U_s=0$ in
Eq.(\ref{Eq:stable_paralleflow_Sun_Int}), this means Galilean
transform for $U$, which make no difference for the following
discussion. If the flow is unstable, it holds
 \begin{equation}
\frac{U(U-U_t)}{\|U-c\|^2}>\frac{1}{p}.
\label{Eq:stable_paralleflow_3rdterm}
 \end{equation}
Otherwise the flow must be stable according to Theorem 1.  From
Eq.(\ref{Eq:stable_paralleflow_3rdterm}), there is
\begin{equation}
U^2+(\frac{2c_r}{p}-U_t)U>\frac{c_r^2+c_i^2}{p}.
\end{equation}
Since $U_t$ is arbitrary, so the left of the inequality is
irrespective to $c_r$. We can rewrite the inequality as
\begin{equation}
c_i^2<p(U^2-U_t U)-c_r^2.
\end{equation}
Obviously, $c_i^2$ reach its largest value at $c_r=0$. So there is
$c_r=U_s$, for $U_s$ is zero.

Result 3 is very important for understanding the instabilities of
inviscid flows. Since $c_i^2$ reach its largest value at
$c_r=U_s$, the most unstable disturbances will propagate with
phase speed of $U_s$. If all the disturbances with $c_r=U_s$ are
neutral stable, then  the other disturbances must be neutral
stable too. This conclusion is of essence, and we state it as a
principle.

Principle: The flow is stable, if and only if all the disturbances
with $c_r=U_s$ are neutrally stable.

By this principle, Rayleigh's criterion can be obtained easily.
Since there is no inflection point, there is no disturbance with
$c_r=U_s$ according to Howard's Semicircle Theorem. Then the flow
is stable according to the Principle, which is what Rayleigh's
Theorem states. Theorem 1 obtained above can also derived from
this Principle, given result 1. The criterion for instability can
also derived from this Principle, we state it as a new theorem.

Theorem 2: If the velocity profile is subject to $f(y)<-\mu_1$
everywhere in the flow, the flow is unstable.

We prove this theorem by proving the following result. Result 4:
If the velocity profile is subject to $f(y)<-\mu_1$ everywhere in
the flow, at least one of the disturbance with $c_r=U_s$ is
unstable.

Proof: According to
Eq.(\ref{Eq:stable_parallelflow_Rayleigh_Int}), for wavenumber
$k=0$ and its first eigenfunction $\phi_1$, it holds
\begin{equation} \int_a^b
[\|\phi_1'\|^2+\frac{U''(U-U_s)}{(U-U_s)^2+c_i^2}\|\phi_1\|^2]\,
dy=0.
 \label{Eq:stable_paralleflow_Result4}
\end{equation}
Then $c_i^2$ must larger than 0. Otherwise,
\begin{equation}
\begin{array}{rl}
&\displaystyle\int_a^b(\|\phi_1'\|^2+\frac{U''(U-U_s)}{(U-U_s)^2+c_i^2}\|\phi_1\|^2
dy\\
 =&\displaystyle\int_a^b (\mu_1+\frac{U''}{U-U_s})\|\phi_1\|^2
dy<0,
\end{array}
\end{equation}
which  contradicts Eq.(\ref{Eq:stable_paralleflow_Result4}). So
the result is proved.

Result 4 shows that if $f(y)$ is less than $-\mu_1$ everywhere in
the flow, there must be unstable disturbances, so the mean flow is
unstable for this case. This is a refinement of Tollmien's
\cite{Tollmien1935} and Lin's \cite{LinCCBook1955}, who proved the
similar result subject to the velocity profile $U(y)$ is either
symmetric or monotone.
%

\begin{figure}
  \includegraphics[width=6cm]{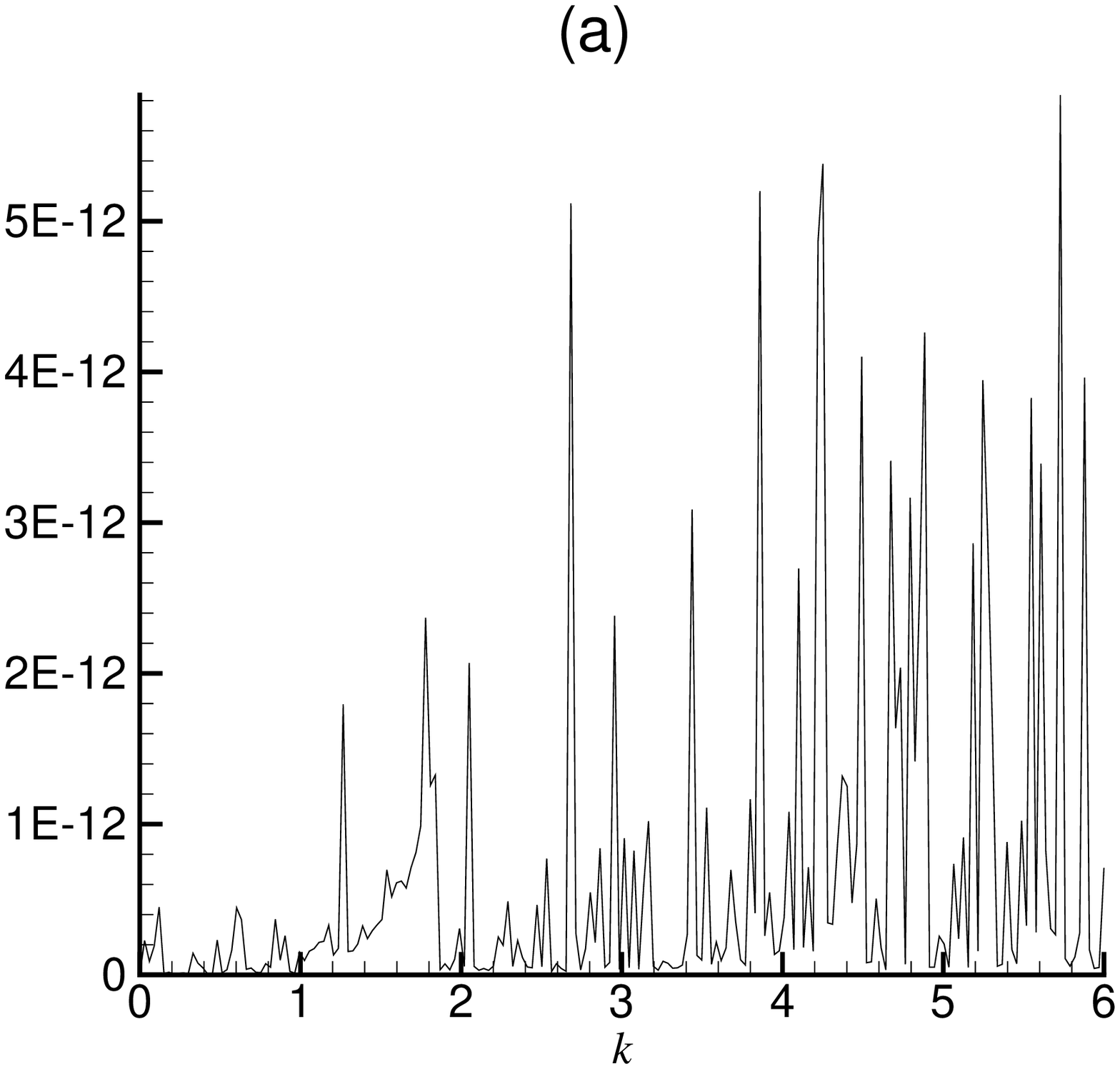}
  \includegraphics[width=6cm]{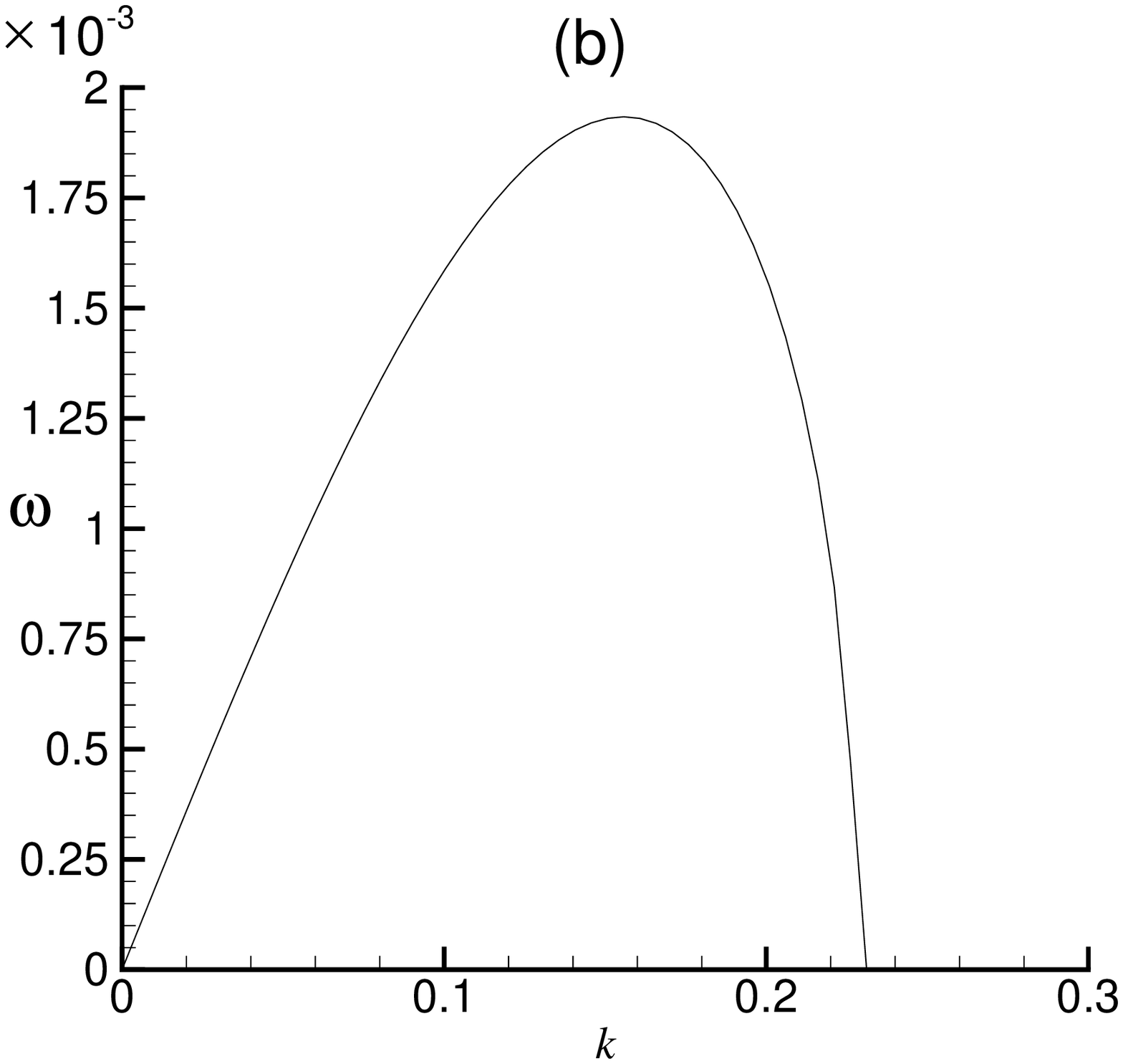}
\caption{Growth rate as an function of wavenumber $k$, (a) for
$U_1=\sin(1.5y)$, (b) for $U_2=\sin(1.6y)$, both within the
interval $-1\leq y\leq 1$. } \label{Fig:Growth}
\end{figure}

To show the power of criterions obtained above, we calculate the
growth rate of two similar velocity profiles $U_1=\sin(1.5y)$ and
$U_2=\sin(1.6y)$ with $-1\leq y\leq 1$. Then there is
$\mu_1=\pi^2/4$ for estimation, and the values of auxiliary
functions are $-2.25$ for $U_1$ and $-2.56$ for $U_2$,
respectively. So $U_1$ is stable according to Theorem 1 and $U_2$
is unstable according to Theorem 2, respectively. In fact, there
are three inflections in the velocity profile $U_2$. The growth
rate of the profiles can be obtained by Chebyshev spectral
collocation method \cite{SchmidBook2000} with 200 collocation
points, as shown in Fig.\ref{Fig:Growth}. It is obvious that
$c_i=0$ for $U_1$ and $c_i>0$ for $U_2$, which agree well with the
theorems obtained above. While Fj\o rtoft's criterion can not
point out the different stability between the two profiles.

On the other hand, Arnold \cite{Arnold1965a,Arnold1969} discussed
the hydrodyanmic stability in a totally different way. He
investigated the conservation law of the inviscid flow and found
two nonlinear stability conditions by means of variational
principle.

Apply Arnold's First Stability Theorem to parallel flow, a stable
criterion is $0<C_1<(U-U_s)/U''<C_2<\infty$ everywhere in the
flow, where $C_1$ and $C_2$ are constants. This corresponds to
Fj\o rtoft's criterion for linear stability, and is well
known\cite{Drazin1981}. Here we find that Theorem 1 proved above
corresponds to Arnold's Second Stability Theorem, i.e., a stable
criterion is $0<C_1<-(U-U_s)/U''<C_2<\infty$ everywhere in the
flow. So It is very interesting that the linear stability
criterions match nonlinear stability theorems very well.

One may note that the criterions of different theorems are
different from each others, $U''(U-U_s)$ for Fj\o rtoft's,
$(U-U_s)/U''$ for Arnold's and $U''/(U-U_s)$ for present works.
But this make no difference for Fj\o rtoft's and Arnold's
criterions in the coarse cases, for example $U''(U-U_s)>0$ is same
as $(U-U_s)/U''>0$. Since the constants $C_1$ and $C_2$ in
Arnold's criterions lack efficiency estimation for applications,
they are not widely used as the linear criterions (eg. Rayleigh's
criterion) be.

An interesting question is which one would be the marginal
criterion, $U''(U-U_s)$ or $U''/(U-U_s)$? It can be seen from
Eq.(\ref{Eq:stable_paralleflow_RayleighEq}) that the stability of
profile $U(y)$ is not only Galilean invariance of $U(y)$, but also
magnitude free of $U(y)$ due to linearity. Since the value of
$U''(U-U_s)$ is only Galilean invariance not magnitude free, it
can not be the the marginal criterion. While the value of
$U''/(U-U_s)$ satisfies both conditions, this is the reason why
the criterions in above theorems are the functions of
$U''/(U-U_s)$.

In summery, the general stability and instability criterions are
obtained for inviscid parallel flow. Those results extend the
former theorems proved by Rayleigh, Tollmien and Fj\o rtoft. The
criterions highlight the vorticity profile for understanding the
instability of the parallel inviscid flow. According to the
criterions, the marginal of instability is limited in a very small
zone. The criterions for stability of parallel inviscid flow can
be applied to barotropic geophysical flow, like Kuo did
\cite{KuoHL1949}. This extension is trivial for the cases of
$f$-plane and $\beta$-plane, and is omitted here. In general,
these criterions will lead future works to investigate the
mechanism of hydrodynamic instability, and shed light on the flow
control and investigation of the vortex dynamics.

The work was original from author's dream of understanding the
mechanism of instability in the year 2000, when the author was a
graduated student and learned the course of hydrodynamics
stability by Prof. Yin X-Y at USTC.

\end{document}